\documentclass[aps,prb,floatfix,reprint,showpacs,10pt]{revtex4-1}

\usepackage{amsfonts}
\usepackage{amssymb}
\usepackage{amsmath}

\usepackage{graphicx}

\begin{document}

\title{Optical conductivity of hydrogenated graphene from first principles}
\author{Sebastian Putz, Martin Gmitra, Jaroslav Fabian}
\affiliation{Institute for Theoretical Physics, University of Regensburg, D-93040 Regensburg, Germany}

\begin{abstract}
We investigate the effect of hydrogen coverage on the optical conductivity of single-side hydrogenated graphene from first principles calculations. To account for different degrees of uniform hydrogen coverage we calculate the complex optical conductivity for graphene supercells of various sizes, each containing a single additional hydrogen atom. We use the linearized augmented plane wave (LAPW) method, as implemented in the WIEN2k density functional theory code, to show that the hydrogen coverage strongly influences the complex optical conductivity and thus the optical properties, such as absorption, of hydrogenated graphene. We find that the optical conductivity of graphene in the infrared, visible, and ultraviolet range has different characteristic features depending on the degree of hydrogen coverage. This opens up new possibilities to tailor the optical properties of graphene by reversible hydrogenation, and to determine the hydrogen coverage of hydrogenated graphene samples in the experiment by contact-free optical absorption measurements.
\end{abstract}


\maketitle

\section{Introduction}
\label{introduction}

Graphene\cite{Novoselov2004, Geim2007} has shaped nanoscience and materials research over the last decade like hardly any other material. Its exceptional electronic structure,\cite{CastroNeto2009} with charge carriers resembling two-dimensional massless Dirac fermions, entails a variety of remarkable properties likely to be harnessed in novel nanoelectronic devices. Adding to the unusual electronic,\cite{Katsnelson2006} mechanical,\cite{Lee2008} and transport\cite{Novoselov2005} properties originating in the linear energy-momentum dispersion of its charge carriers, its optical properties put graphene in the spotlight of optoelectronics and photonics research.\cite{Xia2009, Bonaccorso2010, Mueller2010, Konstantatos2012}

With graphene being a zero-gap semiconductor, a lot of effort has gone into investigating how its band gap can be tuned in a controlled way to combine the advantages of graphene and modern semiconductor devices.\cite{Schwierz2010} Size restriction\cite{Han2007, Li2008} or special substrates,\cite{Giovannetti2007} for example, can introduce a band gap in graphene, but these approaches are irreversible, difficult to implement in devices, or result in fragile band gaps. Chemical functionalization of graphene,\cite{Balog2010, Schedin2007, Gierz2008, Boukhvalov2009, Wehling2009} on the other hand, was demonstrated to be a reversible method to induce robust, tunable band gaps.

Decoration with adatoms like oxygen,\cite{Hossain2012} fluorine,\cite{Robinson2010, Sahin2011} or hydrogen,\cite{Duplock2004, Boukhvalov2008, Ryu2008, Haberer2010, Luo2010} significantly alters the properties of pristine graphene and, in some cases, causes a transition to another class of material altogether: Full hydrogenation of graphene (one completely covered carbon sublattice on each side) leads to the non-magnetic, direct wide-gap semiconductor \textit{graphane}, which was predicted in 2007 by Sofo \textit{et al.}\cite{Sofo2007} from first principles calculations, and demonstrated in the laboratory by Elias \textit{et al.}\cite{Elias2009} two years later. In contrast, as predicted by Zhou \textit{et al.},\cite{Zhou2009} semi-hydrogenation (one completely covered carbon sublattice on one side) produces the ferromagnetic, indirect narrow-gap semiconductor \textit{graphone}. Although the latter system has not yet been synthesized, the transition from graphene to graphone and graphane with increasing degree of hydrogenation shows that the amount of adatom coverage is decisive for the properties of the resulting graphene derivate.

One such property is the presence of magnetic moments, which is particularly important for graphene spintronics.\cite{Zutic2004, Fabian2007, Gmitra2013} Several studies suggest that hydrogenated graphene is indeed magnetic for certain degrees of hydrogenation.\cite{Yazyev2007, Yazyev2008, Yazyev2010, Wang2009, McCreary2012} Optical spectra might thus present an effective means of studying the exchange-split electronic band structure of magnetic hydrogenated graphene with respect to its hydrogen coverage, and enable us to determine if the ground state of hydrogenated graphene is magnetic or not.

The optical conductivity is another property investigated both theoretically and experimentally for many graphene-based systems such as single\cite{Gusynin2007, Dawlaty2008, Mak2008, Stauber2008, Peres2010} and few layer graphene,\cite{Nicol2008, Crassee2011} graphite,\cite{Kuzmenko2008, Trevisanutto2010, Falkovsky2011} and carbon nanotubes.\cite{Ugawa2001}

In this work, we study the influence of hydrogenation on the optical conductivity spectrum of graphene from first principles density functional theory (DFT)\cite{Hohenberg1964, Kohn1965} calculations. We consider three hydrogenated graphene systems with different degrees of uniform single-side hydrogenation, 50\,\%, 12.5\,\%, and 2\,\%, and compare their calculated optical conductivity spectra to the one of pure graphene.

Our results show that the characteristic features of the spectra vary strongly with the degree of hydrogenation, suggesting that the latter could be measured by purely optical---and thus contact-free---methods, and that reversible hydrogenation could be used to tailor the optical properties of graphene in the infrared, visible, and ultraviolet region of the electromagnetic spectrum.

In the following section we present the methods used to obtain the results discussed in Section~\ref{results}, which include the calculated electronic structure and total density of states of various single-side hydrogenated (SSH) graphene systems, as well as their optical conductivity spectra and an analysis of how these are influenced by structural characteristics and the presence of magnetic moments. The summary in Section~\ref{conclusions} concludes this work.

\section{Method}
\label{method}

\begin{figure}
 \centering
 \includegraphics[width=\linewidth]{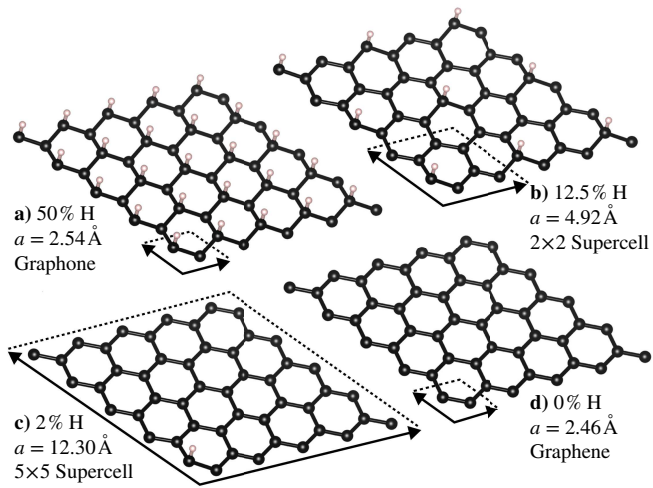}
 \caption{Comparison of different systems investigated in this study. The unit cell of graphone (a) contains 2\,C + 1\,H atoms and has a single-side hydrogen coverage of 50\,\%. The larger 2$\times$2 (b) and 5$\times$5 (c) supercells contain 8\,C + 1\,H and 50\,C + 1\,H atoms, respectively, accounting for 12.5\,\% and 2\,\% of hydrogen coverage. Pure graphene (d) with 2\,C atoms per unit cell corresponds to 0\,\% hydrogen coverage. The lattice constant $a$ is given for each system.}
 \label{Structures}
\end{figure}

Graphene supercells of different size, each containing a single additional hydrogen atom, are used to represent different degrees of single-side hydrogenation (see Fig.\,\ref{Structures}). In the case of 50\,\% SSH graphene (in other words, \textit{graphone}), the modified standard unit cell of graphene contains 2\,C + 1\,H atoms. The 12.5\,\% and 2\,\% SSH graphene systems are modeled by a 2$\times$2 and a 5$\times$5 supercell consisting of 8\,C + 1\,H and 50\,C + 1\,H atoms, respectively. For pure graphene the standard unit cell with two carbon atoms is used. In each system the graphene layers are separated by the vertical unit cell edge length of $ c = 15$\,\AA{} to suppress interlayer coupling.

In the first step, the atomic positions in the three SSH graphene cells are optimized using the plane wave pseudopotential code \textsc{Quantum Espresso},\cite{Giannozzi2009} which implements a quasi-newton algorithm\cite{Broyden1970, Fletcher1970, Goldfarb1970, Shanno1970} for atomic force relaxation (see Tab.\,\ref{StructuralParameters} for an overview of the structural parameters after relaxation). We use ultra-soft pseudopotentials\cite{Vanderbilt1990} for carbon and hydrogen, and the Perdew-Burke-Ernzerhof variant of the generalized gradient approximation (PBE-GGA)\cite{Perdew1996} for the exchange-correlation functional. The plane wave basis set is independent of the atom positions and species, which is why the calculated total forces on the atoms are true Hellman-Feynman\cite{Feynman1939} forces, without the need for basis-set corrections (Pulay forces).\cite{Pulay1969} This, combined with the computational efficiency of the plane wave basis set, makes \textsc{Quantum Espresso} a suitable choice for the initial structural optimization of large supercells (the 5$\times$5 supercell contains 51 atoms). The resulting structures are found to be sufficiently relaxed to obtain well-converged optical properties in the subsequent steps of the calculation.

The second step involves the calculation of the electronic band structure and the total density of states (DOS) of each system with the full-potential linearized augmented plane wave (LAPW)\cite{Andersen1975} code WIEN2k.\cite{Blaha2001} We again choose the PBE-GGA as the exchange-correlation functional and obtain the self-consistent electronic ground state density for each system. Here, the convergence criterion is that the charge distance between two consecutive iterations of the self-consistent field cycle, integrated over the unit cell, be smaller than $10^{-5}e$, where $e$ is the positive value of the elementary charge. In order to study the effect of adatom-induced magnetic moments on the optical properties of hydrogenated graphene, we perform both a non-magnetic calculation (this means the calculation explicitly disregards the electron spin), and a spin-polarized calculation for each system except pure graphene, for which we perform only a non-magnetic reference calculation.

In the last step we use the converged systems of Kohn-Sham\cite{Kohn1965} eigenenergies and eigenstates to obtain the imaginary part of the complex dielectric function in linear response according to\cite{Kubo1956, Kubo1957, AmbroschDraxl2006}

\begin{align}
 \mathrm{Im}\left[\epsilon_{\alpha\beta}(\omega)\right] = & \frac{\hbar^2 e^2}{\pi m_e^2 \omega^2} \sum_{\mathrm{n} \neq \mathrm{n}'} \,\int_{\mathbf{k}} \Pi_{\mathrm{n}\mathrm{n}',\mathbf{k}}^\alpha\,\Pi_{\mathrm{n}'\mathrm{n},\mathbf{k}}^\beta\label{DielectricFunction}\\
&\times \left( f(\epsilon_{\mathrm{n},\mathbf{k}}) - f(\epsilon_{\mathrm{n}',\mathbf{k}}) \right) \,\delta(\epsilon_{\mathrm{n}',\mathbf{k}} - \epsilon_{\mathrm{n},\mathbf{k}} - \hbar\omega). \nonumber
\end{align}

Here, $\Pi_{\mathrm{n}\mathrm{n}',\mathbf{k}}^\alpha = \langle\mathrm{n}',\mathbf{k}|\hat{\mathrm{p}}_\alpha|\mathrm{n},\mathbf{k}\rangle$ is the transition matrix element of the $\alpha$-component of the momentum operator for a direct interband transition ($\mathrm{n} \neq \mathrm{n}'$) from the initial Kohn-Sham state $|\mathrm{n},\mathbf{k}\rangle$ with energy $\epsilon_{\mathrm{n},\mathbf{k}}$ into the final state $|\mathrm{n}',\mathbf{k}\rangle$ with energy $\epsilon_{\mathrm{n}',\mathbf{k}}$, $f(\epsilon_{\mathrm{n},\mathbf{k}})$ is the Fermi-Dirac distribution function evaluated at energy $\epsilon_{\mathrm{n},\mathbf{k}}$, $m_e$ denotes the electron mass, and $\omega$ is the angular frequency of the electromagnetic radiation causing the transition.

The real part of the complex optical conductivity is calculated from the imaginary part of the complex dielectric function of Eq.\,(\ref{DielectricFunction}) using\cite{Dressel2001}

\begin{equation}
 \mathrm{Re}\left[\sigma_{\alpha\beta}(\omega)\right] = \frac{4\pi}{\omega} \mathrm{Im}\left[\epsilon_{\alpha\beta}(\omega)\right].
 \label{EpsilonToSigma}
\end{equation}

However, in our case this results in the optical conductivity for the three-dimensional slab supercell whose edge length $c$ perpendicular to the graphene layer produces an interlayer spacing large enough to prevent any hybridization of states pertaining to adjacent graphene sheets. Multiplying the result of Eq.\,(\ref{EpsilonToSigma}) by the interlayer spacing $ c = 15$\,\AA{} leads to the desired value of the optical conductivity for the essentially two-dimensional hydrogenated graphene film.

To achieve better comparability we normalize the calculated spectra to the universal AC optical conductivity of graphene,\cite{Ando2002, Gusynin2006, Falkovsky2007, Nair2008} given by $\sigma_0 = e^2/(4\hbar)$.

\begin{table}[tb]
 \caption{\label{StructuralParameters} Structural parameters of the investigated hydrogenated graphene systems after relaxation. The in-plane supercell edge length is denoted by $a$ (the vertical edge length $ c = 15$\,\AA{} for each supercell), while the C--H bond length, and the C--C distance between the carbon atoms surrounding the hydrogenated carbon site are given by $d_\mathrm{H}$ and $d_\mathrm{CC}$, respectively. The parameter $\Delta$ describes the vertical distance between the hydrogenated carbon atoms and their neighbors, with the ratio $\Delta/d_\mathrm{CC}$ being a measure for the out-of-plane distortion induced by hydrogenation. All lengths are given in \AA{}ngstr\"oms [\AA].}
\begin{tabular*}{0.9\linewidth}{@{\extracolsep{\fill} }lrccccc}
 System & SSH & $a$ & $d_\mathrm{H}$ & $d_\mathrm{CC}$ & $\Delta$ & $\Delta/d_\mathrm{CC}$\\
\hline
\hline
Graphone & 50\,\% & 2.54 & 1.158 & 2.537 & 0.322 & 12.9\,\%\\
2$\times$2 & 12.5\,\% & 4.92 & 1.131 & 2.500 & 0.344 & 13.8\,\%\\
5$\times$5 & 2\,\% & 12.30 & 1.133 & 2.513 & 0.325 & 12.9\,\%\\
\hline
\end{tabular*}
\end{table}

\section{Results}
\label{results}

\begin{figure}[t]
 \centering
 \includegraphics[width=\linewidth]{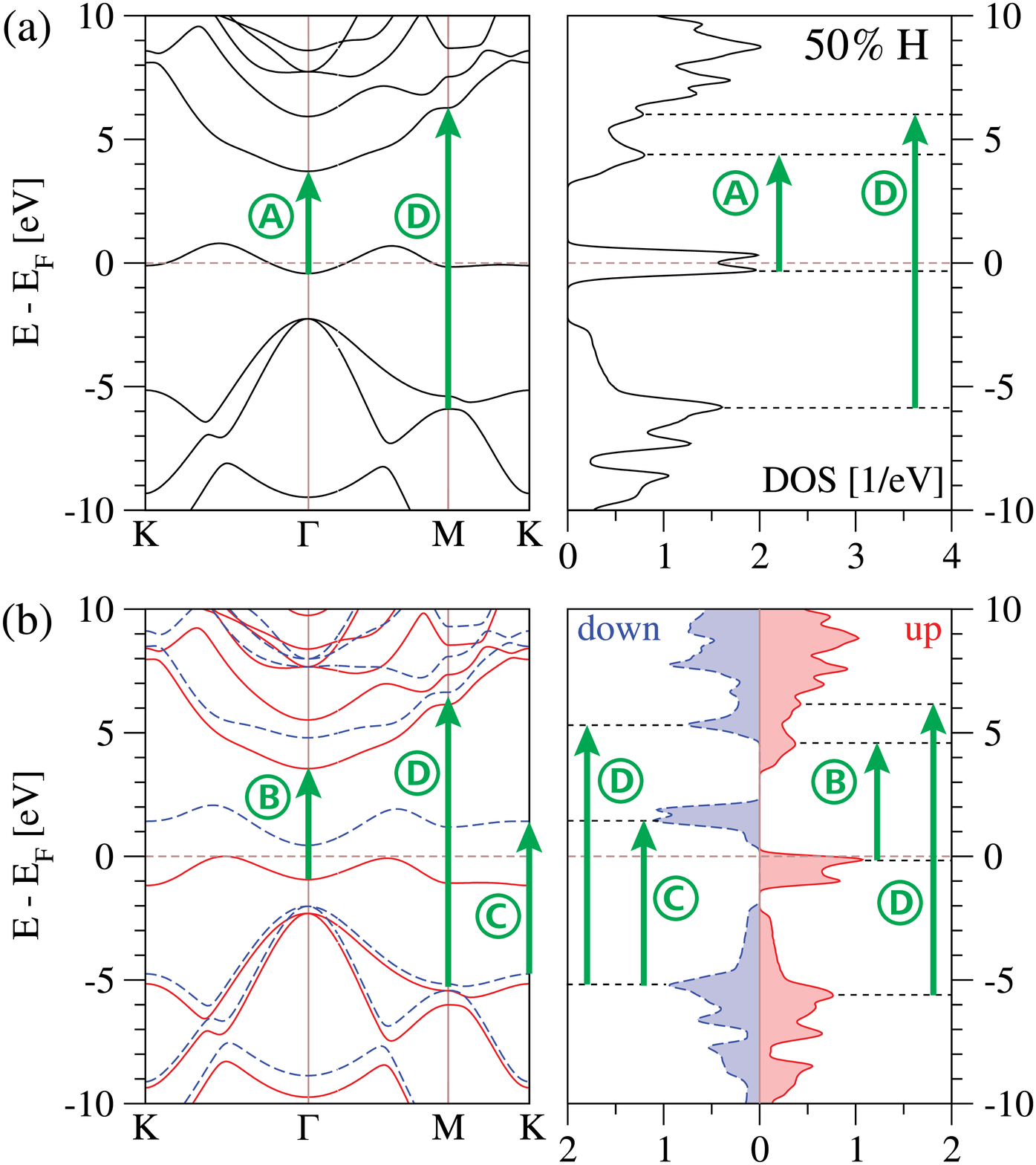}
\caption{(Color online) Calculated electronic band structure along high-symmetry lines in the first Brillouin zone (left panels) and broadened total density of states (DOS) per unit cell (right panels) for the non-magnetic (a) and the spin-polarized (b) 50\,\% SSH case (\textit{graphone}). In the spin-polarized case the spin-resolved total DOS is shown; quantities associated with spin up (down) are shown as solid red (dashed blue) lines. Energies are given relative to the Fermi energy $\mathrm{E}_\mathrm{F}$. Labeled arrows indicate direct interband transitions corresponding to pronounced features in the optical conductivity spectra (see Fig.\,\ref{fig:OptCond}).}
\label{fig:banddos1x1}
\end{figure}

\begin{figure}[t]
 \centering
 \includegraphics[width=\linewidth]{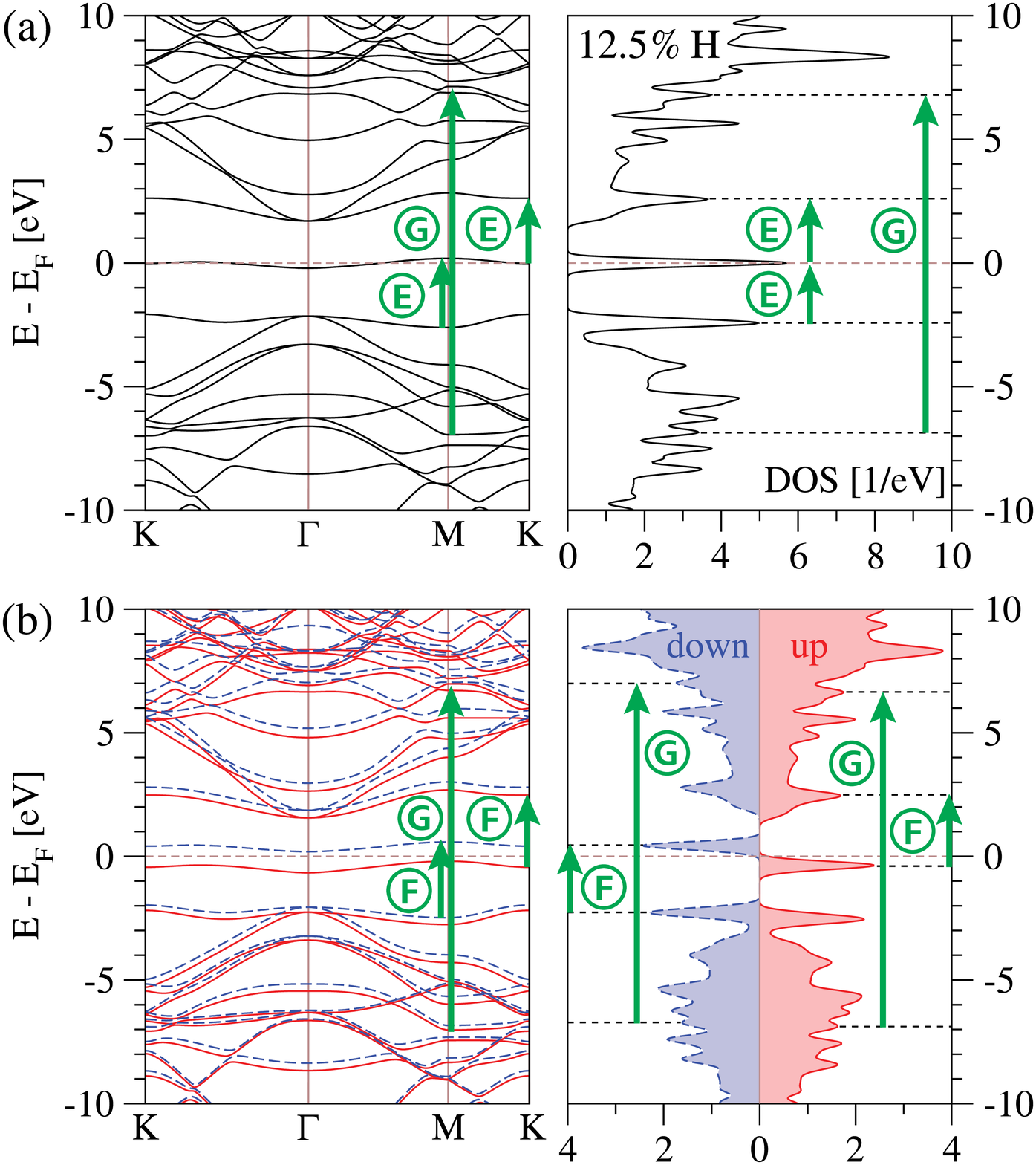}
\caption{(Color online) See the caption of Fig.\,\ref{fig:banddos1x1}, but for the 12.5\,\% SSH case (2$\times$2 supercell).}
\label{fig:banddos2x2}
\end{figure}

\begin{figure}[t]
 \centering
 \includegraphics[width=\linewidth]{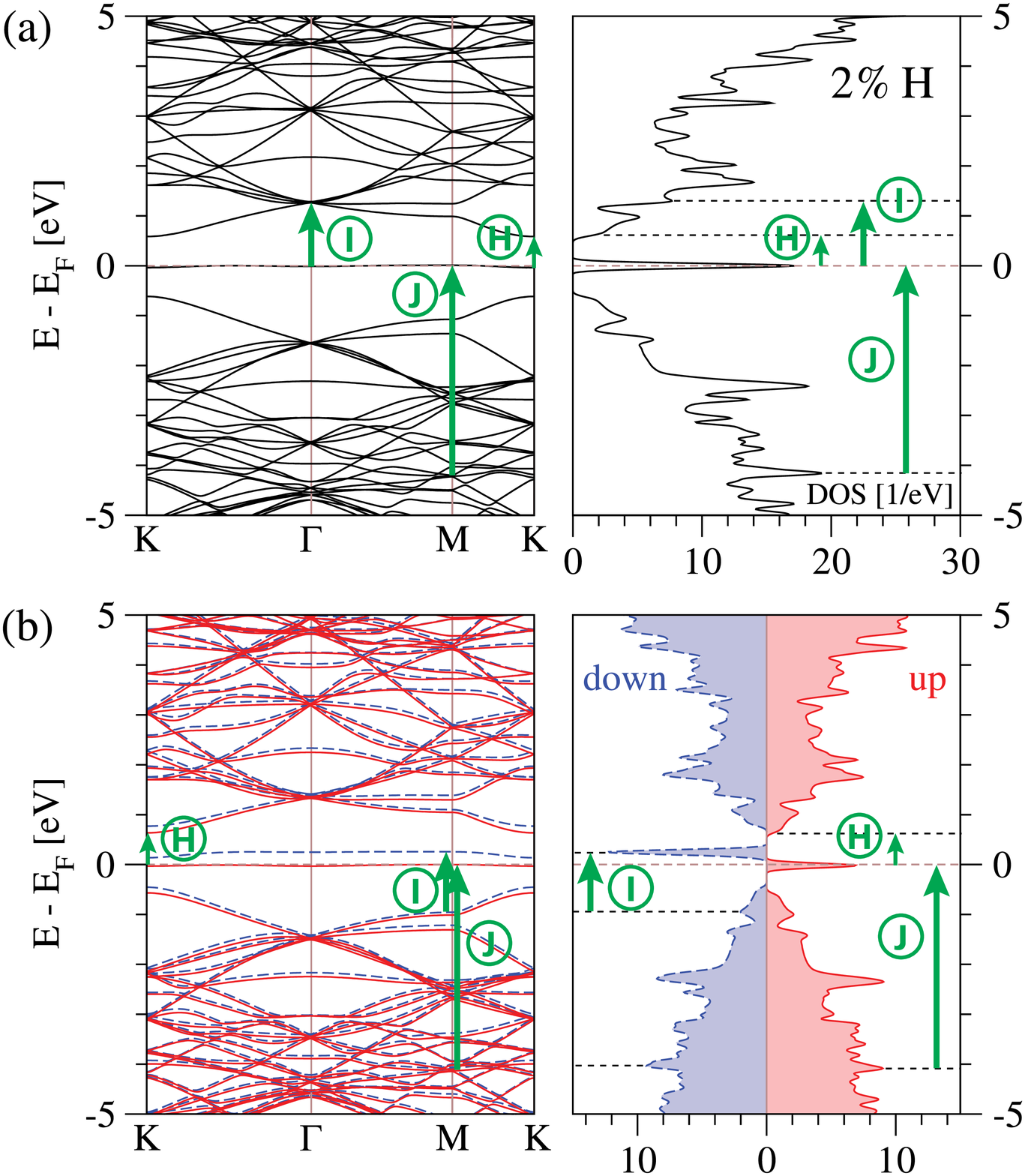}
\caption{(Color online) See the caption of Fig.\,\ref{fig:banddos1x1}, but for the 2\,\% SSH case (5$\times$5 supercell).}
\label{fig:banddos5x5}
\end{figure}

\begin{figure}[t]
 \centering
 \includegraphics[width=\linewidth]{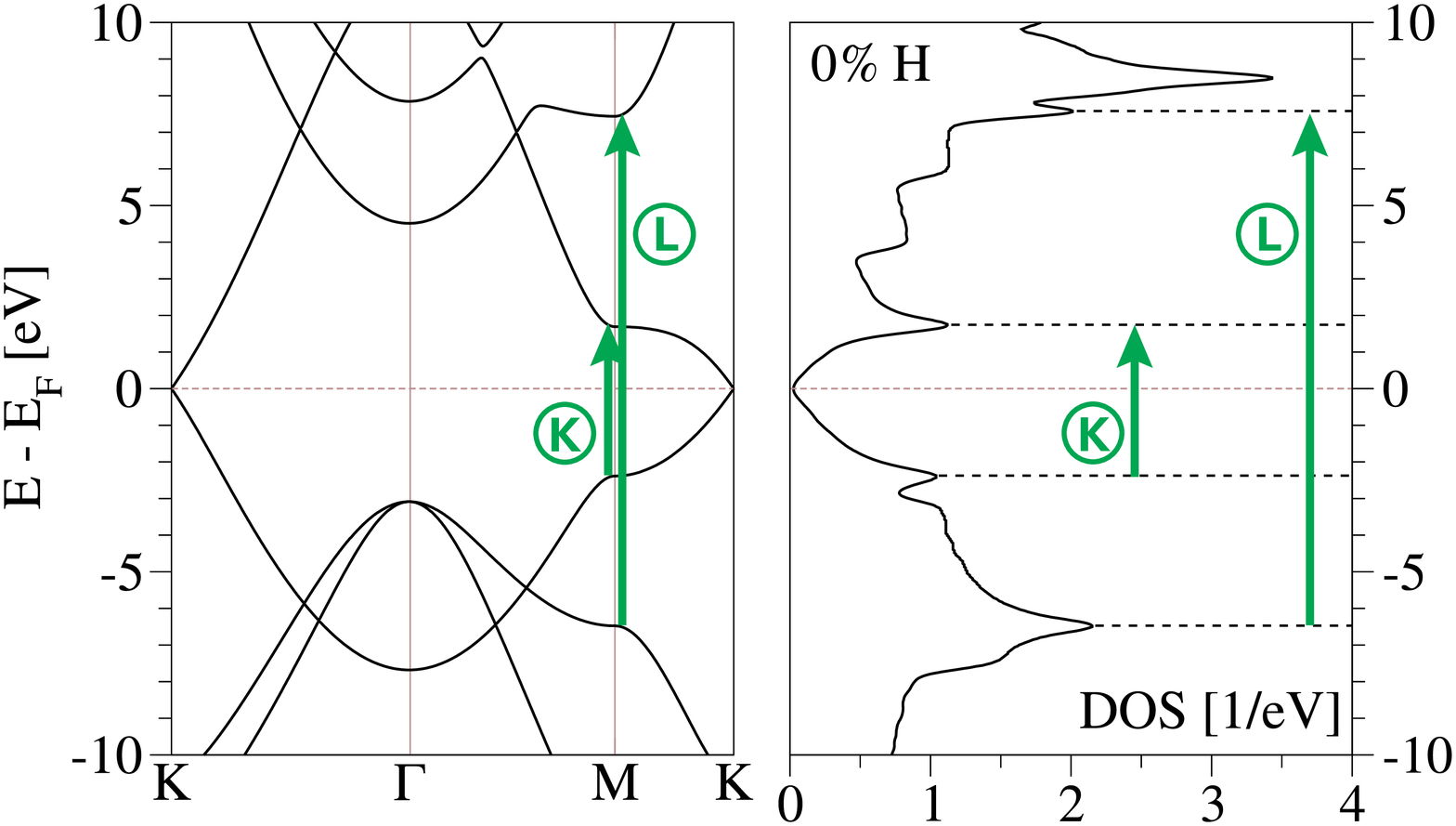}
\caption{(Color online) See the caption of Fig.\,\ref{fig:banddos1x1}, but for non-magnetic pure graphene.}
\label{fig:banddosPureGrp}
\end{figure}

\begin{table}[tb]
 \caption{\label{BandGaps} Calculated band gaps at the $\Gamma$ and K high-symmetry points (non-magnetic case), and exchange splitting at the Fermi energy (spin-polarized case) for each SSH graphene system. All energy difference values are given in electron volts [eV].}
\begin{tabular*}{0.9\linewidth}{@{\extracolsep{\fill} }lrccc}
 System & SSH & $\Gamma$-gap & K-gap & exchange splitting\\
\hline
\hline
Graphone & 50\,\% & 3.93 & 8.20 & 2.01\\
2$\times$2 & 12.5\,\% & 1.87 & 2.63 & 0.80\\
5$\times$5 & 2\,\% & 1.27 & 0.62 & 0.27\\
\hline
\end{tabular*}
\end{table}

While each carbon atom in pure graphene is covalently bonded to its three neighbors, thus being sp\textsuperscript{2}-hybridized, hydrogenated carbon atoms in hydrogenated graphene are closer to an sp\textsuperscript{3}-hybridized tetrahedral conformation. This results in a buckling of the graphene sheet in the vicinity of hydrogenated carbon sites, as the carbon atoms carrying a hydrogen atom are shifted out-of-plane to adopt an energetically more favorable tetrahedral conformation (see Tab.\,\ref{StructuralParameters}).\cite{Duplock2004} In the dilute hydrogenation limit, a single, isolated hydrogen adatom affects the properties of graphene only locally, whereas a dense hydrogen coverage profoundly impacts its atomic and electronic structure. For example, each additional isolated hydrogen adatom adds a magnetic moment of about $1\,\mu_B$ (Bohr magneton) to the system.\cite{Yazyev2007}

For each hydrogenated graphene system we thus perform a spin-polarized and---to study the influence of magnetic moments on the electronic band structure and the optical conductivity---a non-spin-polarized calculation, with the latter resulting in a non-magnetic system. We use the calculated electronic band structure and total DOS presented in Figs.\,\ref{fig:banddos1x1}--\ref{fig:banddosPureGrp} to identify those direct interband transitions that significantly contribute to the pronounced features of the optical conductivity spectra shown in Fig.\,\ref{fig:OptCond}. The transitions indicated by arrows and Roman capital letters in Figs.\,\ref{fig:banddos1x1}--\ref{fig:banddosPureGrp} serve as representatives for all transitions that can occur between a given pair of bands at different values of $\mathbf{k}$, and match the features marked with the same capital letters in Fig.\,\ref{fig:OptCond}. Table\,\ref{BandGaps} provides a summary of the calculated values for band gaps at high-symmetry points and the exchange splittings at the Fermi energy. The latter have been derived by determining the distance between corresponding characteristic peaks in the spin-resolved total densities of states (see bottom right panels in Figs.\,\ref{fig:banddos1x1}--\ref{fig:banddos5x5}).

\subsection{Electronic Structure}
\label{electronicstructure}

In the dense limit of 50\,\% SSH graphene (\textit{graphone}; see Fig.\,\ref{fig:banddos1x1}), in the spin-polarized as well as the non-magnetic case, the Dirac cone at the K-point is completely absent. In fact, there are no conic features whatsoever. In both cases band gaps open, and a relatively flat midgap state (which is exchange-split in the spin-polarized case) appears close to the Fermi energy. While this midgap state crosses the Fermi energy in the non-magnetic case, resulting in a metallic band structure, its exchange-split equivalent in the spin-polarized case leads to an indirect band gap of about $0.47\,\mathrm{eV}$ between the valence band maximum along the $\overline{\mathrm{K}\Gamma}$ high-symmetry line and the conduction band minimum at the $\Gamma$-point. Having a high density of states, these bands are responsible for the characteristic features of the optical conductivity spectra at energies $\lessapprox 5\,\mathrm{eV}$ because they provide the initial or final states for many transitions.

An intermediate single-side hydrogenation value of 12.5\,\% (see Fig.\,\ref{fig:banddos2x2}) presents a band structure similar to the previous case, but with a smaller exchange splitting of the bands in the spin-polarized calculation (see Tab.\,\ref{BandGaps}). The midgap states become more flat and their smaller splitting results in a smaller indirect band gap of about $0.39\,\mathrm{eV}$ in the spin-polarized case, whereas the non-magnetic case is metallic. This is consistent with the smaller areal density of magnetic moments in the 2$\times$2 supercell as compared to the 1$\times$1 cell of \textit{graphone}, illustrating the decreased influence of the magnetic moments in this case of medium hydrogen coverage. Furthermore, the states at the K-point of the neighboring bands (or exchange-split band pairs) above and below the midgap states are closer to the Fermi level than in the 50\,\% SSH case.

The exchange splitting in the dilute limit of 2\,\% SSH graphene (see Fig.\,\ref{fig:banddos5x5}) is even smaller than for the previous systems (see Tab.\,\ref{BandGaps}), aligning the spin-polarized with the non-magnetic band structure. This is because the additional magnetic moments introduced by the hydrogen adatoms are distributed over 50 carbon atoms of the 5$\times$5 supercell. The midgap states are almost completely flat, and the neighboring bands above and below the midgap states approach each other at the K-points, eventually reforming a Dirac cone when the amount of hydrogen coverage is further reduced below 2\,\%.

Pure graphene (corresponding to 0\,\% SSH; see Fig.\,\ref{fig:banddosPureGrp}) serves as a reference calculation. The system is a zero-gap semiconductor characterized by the linear dispersion relation (Dirac cones) in the vicinity of the K-points.

\subsection{Optical Conductivity}
\label{opticalconductivity}

\begin{figure}[t]
 \centering
 \includegraphics[width=\linewidth]{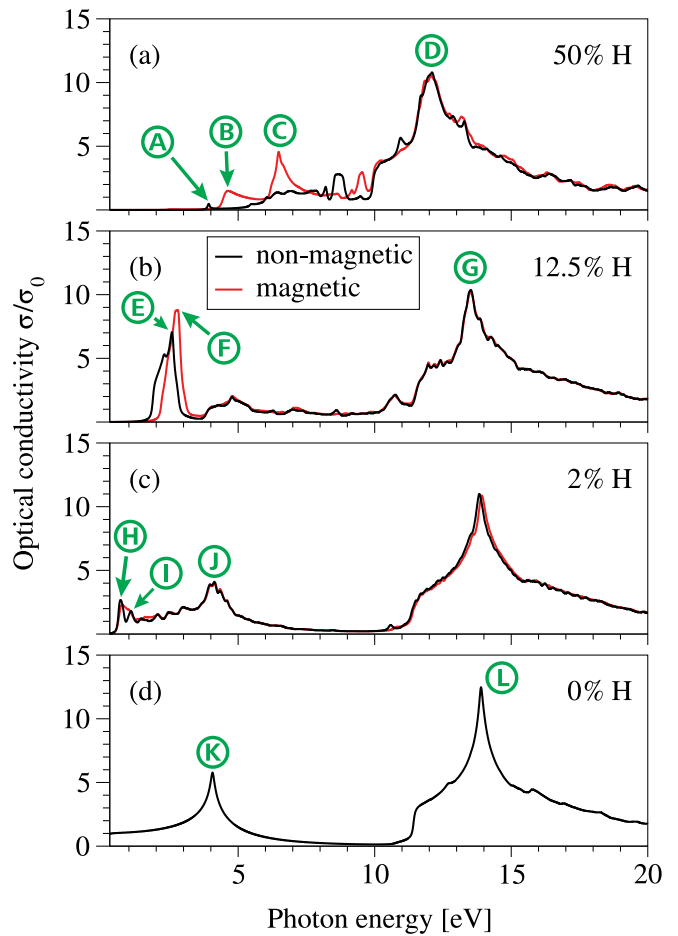}
 \caption{(Color online) Calculated real part of the complex optical conductivity $\sigma$ for 50\,\% (a), 12.5\,\% (b), and 2\,\% (c) SSH graphene, as well as pure graphene (d), given in units of the universal AC optical conductivity $\sigma_0$ of graphene. Pronounced features of the spectra are labeled in concordance with the arrows in Figs.\,\ref{fig:banddos1x1}--\ref{fig:banddosPureGrp}, indicating the most important transitions contributing to them.}
 \label{fig:OptCond}
\end{figure}

The optical conductivity spectra for all four systems (see Fig.\,\ref{fig:OptCond}) are obtained from the electronic band structure results using the method described in Section~\ref{method}. The spectra are calculated for photon energies ranging from 0.3\,eV to 20\,eV with a resolution of 1.36\,meV, comprising the infrared, visible, and ultraviolet (IR-VIS-UV) parts of the electromagnetic spectrum. Lorentzian broadening of 50\,meV is applied to account for finite-lifetime effects.

For each hydrogenated system the results for both the spin-polarized and the non-magnetic calculation are shown. In the following, capital letters in parentheses, such as (A), refer to the labels used in Figs.\,\ref{fig:banddos1x1}--\ref{fig:OptCond}. Each paragraph deals with one of the three investigated hydrogenated graphene systems.

Within the calculated spectral range, \textit{graphone} (50\,\% SSH graphene; see Fig.\,\ref{fig:OptCond}a) is transparent for photon energies below 5\,eV, except for a small peak (A) at 3.9\,eV. The most prominent feature is a broad peak around 12.1\,eV (D). For energies between 4\,eV and 10\,eV the spectra of the non-magnetic and the spin-polarized case differ considerably as the spin-split band structure of the latter gives rise to two new peaks at 4.6\,eV (B) and 6.6\,eV (C), while the spectra are similar for energies above 10\,eV. This indicates that for 50\,\% hydrogenation the spectrum is significantly influenced by magnetic moments for photon energies below 10\,eV.

An interesting phenomenon occurs in 12.5\,\% SSH graphene (see Fig.\,\ref{fig:OptCond}b). Coincidentally, the transitions to and from the midgap state (or the exchange-split midgap states in the spin-polarized case) are of the same energy, and both contribute to a pronounced peak at 2.5\,eV (E) or 2.8\,eV (F) in the visible part of the electromagnetic spectrum between 1.5\,eV and 3\,eV. The center of the broad peak similar to the one in the 50\,\% SSH case is shifted to higher energies and centered at 13.6\,eV (G).

The dilute 2\,\% hydrogenation case (see Fig.\,\ref{fig:OptCond}c) shows many features in the low-energy region from 0.3\,eV to 5\,eV, the most important of which are the absorption peaks at 0.7\,eV (H), 1.1\,eV (I), and 4.2\,eV (J). The broad peak centered at 13.9\,eV is shifted to higher energies compared to the previous two cases. The spectra for the non-magnetic and the spin-polarized case hardly differ, which is consistent with the low areal density of magnetic moments in the 2\,\% SSH case (1\,$\mu_B$ per 50 carbon atoms). The overall shape of the spectrum is approaching the reference spectrum of pure graphene (see Fig.\,\ref{fig:OptCond}d), whose most pronounced features are the peaks at 4.1\,eV (K) and 13.9\,eV (L).

\subsection{Influence of Structure and Magnetic Moments}
\label{comparison}

\begin{figure}[t]
 \centering
 \includegraphics[width=\linewidth]{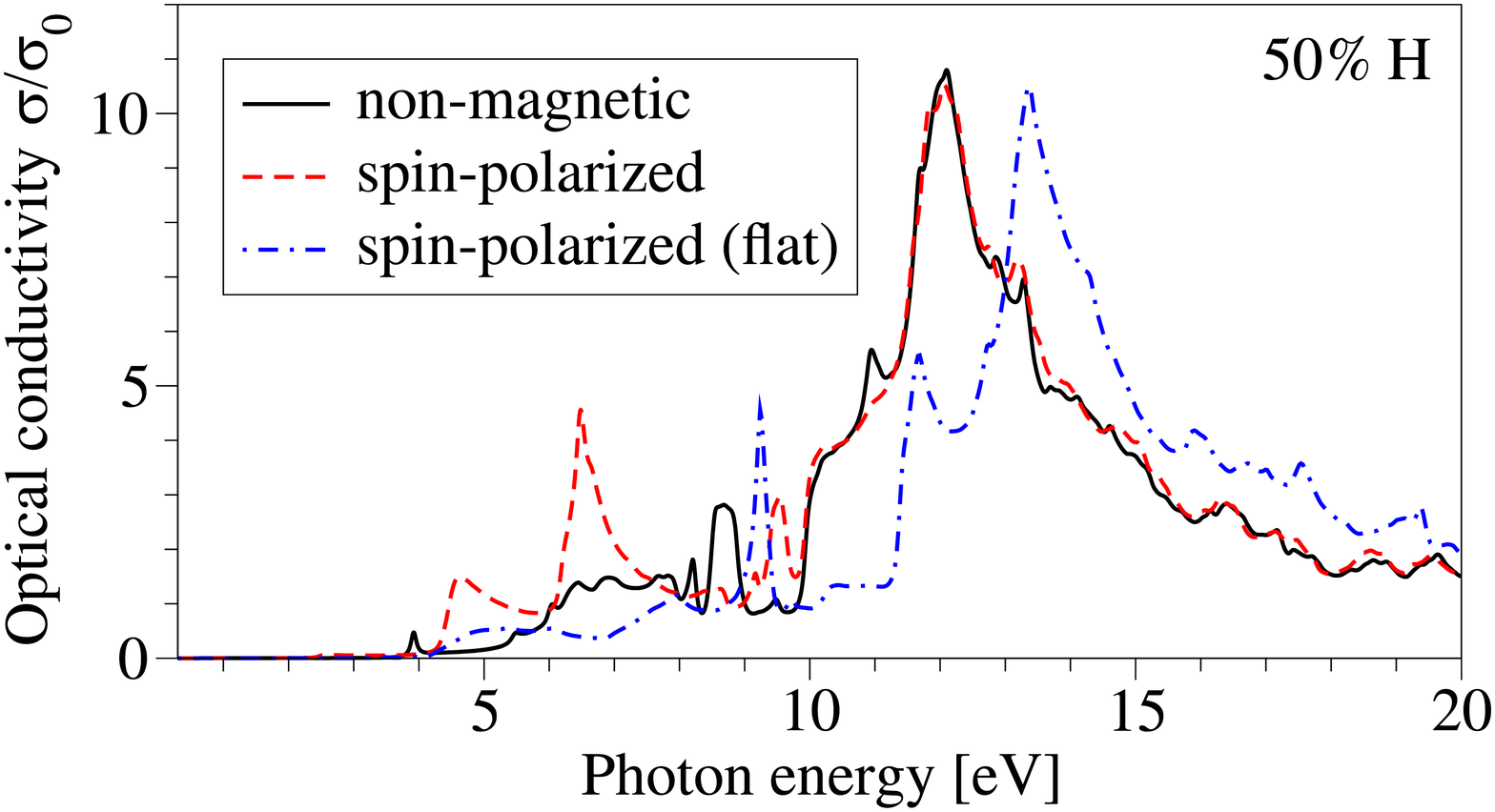}
 \caption{(Color online) Calculated real part of the complex optical conductivity $\sigma$ in units of the universal AC optical conductivity $\sigma_0$ of graphene for 50\,\% SSH graphene, comparing the non-magnetic, the spin-polarized, and another spin-polarized but unrelaxed flat case, in which all carbon atoms are restricted to the same plane.}
 \label{fig:1x1comparison}
\end{figure}

In order to determine how strongly the presence of magnetic moments, or of different structural characteristics, influences the optical conductivity spectrum of hydrogenated graphene, we compare the spectra of 50\,\% SSH graphene for the following three cases: the non-magnetic case, the spin-polarized case, and an artificially flat spin-polarized case in which all carbon atoms are restricted to the same plane. The results are shown in Fig.\,\ref{fig:1x1comparison}, which is equivalent to Fig.\,\ref{fig:OptCond}a, except for the additional curve of the flat spin-polarized case.

The spectrum of the flat spin-polarized case is shifted towards higher energies with respect to the non-magnetic case, but the overall shape is similar and both cases have their most pronounced spectral features in common. In contrast, the result for the relaxed (buckled) spin-polarized case is very different (as discussed in Section~\ref{opticalconductivity}). This indicates that within our PBE-GGA calculation magnetism in the flat case is quenched. However, Casolo \textit{et al.}\cite{Casolo2010} demonstrated that the quenching of magnetism in the flat hydrogenated graphene system is due to the self-interaction error afflicting GGA functionals such as PBE, and thus not physical, and that magnetism in the flat system is recovered in calculations employing hybrid functionals such as PBE0, which mixes PBE exchange with Hartree-Fock exchange. Hence, we conclude that the influence of structural characteristics on the optical properties of hydrogenated graphene is secondary. The influence of magnetic moments, if present, is considerably larger, which is why we expect an optical measurement to be able to detect the presence of magnetic moments in real hydrogenated graphene samples.

\section{Conclusions}
\label{conclusions}

We studied the influence of hydrogenation on the optical conductivity of hydrogenated graphene from first principles calculations. Different degrees of hydrogenation were simulated by optimized-geometry graphene supercells of different size, each containing an additional hydrogen atom. Performing both an explicitly non-magnetic and a spin-polarized calculation for each supercell, we obtained the electronic band structure and total density of states for 50\,\%, 12.5\,\%, 2\,\%, and 0\,\% hydrogenated graphene. These results were used to calculate the corresponding optical conductivity spectra in linear response over the IR-VIS-UV range of the electromagnetic spectrum.

While the dense hydrogenation in the 50\,\% SSH case exhibited a spectrum distinct from the one of pure graphene, the influence of the local tetrahedral conformation of the hydrogenated carbon atoms and the resulting magnetic moments degraded with decreasing hydrogenation density. For intermediate values of hydrogenation we observed the coincidental appearance of a pronounced peak in the optical conductivity in the visible part of the spectrum.

Since the influence of hydrogenation on the optical conductivity was found to be significant, one could employ optical measurement techniques (for example an absorption measurement) to monitor the hydrogenation process, or one could tailor the optical conductivity of graphene by reversible hydrogenation. Furthermore, our results suggest that an optical measurement could determine if the ground state of hydrogen-functionalized graphene is magnetic or not.

Finally, a comparison of three different calculations of the 50\,\% SSH case showed that, within the PBE generalized gradient approximation to the exchange-correlation functional, structural changes induced by hydrogen adatoms are ultimately responsible for additional magnetic moments and hydrogenation-dependent optical conductivity spectra.

Upon completion of our manuscript we larned about the work of Cheng \textit{et al.},\cite{Cheng2013} which also deals with first principles optical spectra of hydrogenated graphene. Our numerical data of the diagonal components of the optical conductivity agree well with those of Cheng \textit{et al.} Unfortunately, we cannot confirm the results for the off-diagonal component $\sigma_{xy}$, which quantifies effects such as Faraday rotation. Even at the very high degree of numerical precision of 15,760 k-points in the irreducible Brillouin zone for the optics calculation, the convergence of $\sigma_{xy}$ (which is numerically very subtle to calculate) eluded us for the small 50\,\% SSH graphene supercell, even more so for the 2$\times$2 and 5$\times$5 supercells. This is why we do not present such data here. Our pre-convergence results for $\sigma_{xy}$ are not only one order of magnitude smaller than those of Cheng \textit{et al.}, they also do not exhibit any similar
trend. A possible reason is that the method of Cheng \textit{et al.} relies on an interpolation technique for the k-point grid in the irreducible Brillouin zone. Our method does not make such an approximation.

Going beyond Kohn-Sham DFT with a GW calculation, the electronic structure of hydrogenated graphene exhibits larger band gaps,\cite{Lebegue2009, Fiori2010, Kharche2011} leading to a shift in energy of the characteristic optical conductivity peaks of the spectra presented in this work. Although quantitatively different, the linear response spectra derived from a GW calculation should remain qualitatively unchanged. We thus expect our main results to be valid beyond standard DFT.\\

\acknowledgments

This work was supported by GRK 1570 of the German Research Foundation.

\newpage

\bibliography{GraphenePaper}

\end{document}